\documentclass{article}
\usepackage{hiph-preprint}
\usepackage{graphicx}
\usepackage{subfigure}
\volnumber{22} \issuenumber{1} \edyear{2005}                             
\frompage{000} \topage{000}                                              
\recrevdate{31 October 2005}                                              
\def\rightmark{Hubble-like flows ... }
\def\leftmark{M.Chojnacki}
\title{Hubble-like flows in relativistic heavy-ion collisions} 
\authors{
  {Mikolaj Chojnacki \index{Chojnacki, Mikolaj}} \\[2.812mm]
  {\normalsize \hspace*{-8pt} Institute of Nuclear Physics, Polish Academy of Sciences, 31-342 Krak\'ow, Poland\\[0.2ex]}
}
\abstract{We study the dynamical appearance of scaling solutions in relativistic hydrodynamics. The phase transition effects are included through the temperature dependent sound velocity. If a pre-equilibrium transverse flow is included in the initial conditions, then it may reach the form of the asymptotic Hubble flow, $r/t$, in short evolution times, 7-15 fm. The numerical solutions are found to support to the freeze-out models (Blast-Wave, Buda-Lund, Cracow). }
\makeindex
\begin{document}
\maketitle
\vspace*{-0.5cm}
\section{Introduction}\label{intro}
This article is a contribution to the Quark Matter 2005 poster proceedings, prepared with W. Florkowski and T. Cs\"org\H{o}, and based on our recent paper \cite{Chojnacki:2004ec}.\par
The successful parameterizations \cite{Schnedermann:1993ws,Csorgo:1999sj,Broniowski:2001we} of the freeze-out process at RHIC indicate that the system formed in Au+Au collisions is highly thermalized and exhibits strong longitudinal and transverse expansion that may be characterized by the Hubble law, $\mathbf{v}=H\mathbf{r}$.
Such form of expansion is used in cosmology, with $H$ being a function of time (in spite of the time dependence, $H$ is commonly called the Hubble constant). The analogy between the Hubble law used in cosmology and in relativistic hydrodynamics is related to the dependence of the four-velocity $u^\mu$ on the space-time position of the fluid element $x^\mu$,
\begin{equation}
u^\mu=\frac{x^\mu}{\tau}=\frac{t}{\tau}\left(1,\frac{x}{t},\frac{y}{t},\frac{z}{t}\right).
\label{4vel}
\end{equation} 
Here $\tau =\sqrt{t^2-x^2-y^2-z^2}$ is the invariant time. The form of the four-velocity displayed by Eq.  (\ref{4vel}) is known as the {\itshape scaling solution}. In many cases it represents the asymptotic solution of the hydrodynamic equations \cite{Landau:1953gs,Cooper:1975qi,Ollitrault:1992bk}.
In our approach we assume that the elementary parton-parton collisions that lead to thermalization of the system also lead to collective behavior, i.e., to the development of the pre-equilibrium transverse flow. This type of the flow has to be included in the initial conditions for the hydrodynamic equations. This is achieved by the use of the following transverse-flow profile at the initial time $t_0 = 1 \hbox{fm}$, 
\begin{equation}
v^{0}_r = v_r(r,t=t_0) = \frac{H_0 r}{\sqrt{1 + H_0^2 r^2}}.
\label{initvr}
\end{equation}
The quantity $H_0$ may be interpreted here as the Hubble constant that controls the amount of the pre-equilibrium flow. In the general case, $H_0$ differs from $H$ which is determined by the flow profile at much later times.
\section{Hydrodynamic equations}\label{hydroeqn}
Here we briefly introduce the hydrodynamic equations in the way they were formulated in Ref. \cite{Baym:1983sr}. The conservation of energy and momentum, $\partial_\mu T^{\mu\nu}=0$, in the case of an ideal fluid with zero net baryon density has the following form
\begin{equation}
u^{\mu }\partial _{\mu }\left( T\,u^{\nu }\right) = \partial ^{\nu }T, \quad\quad 
\partial _{\mu }\left( \sigma u^{\mu }\right) = 0, 
\label{hydTs}
\end{equation}
where $T$ is the temperature, $\sigma $ is the entropy density, and $u^{\mu}=\gamma \left( 1,\mathbf{v}\right) $ is the hydrodynamic four-velocity (with $\gamma =1/\sqrt{1-v^{2}}$). The role of equation of state is played by the temperature dependent sound velocity. 
\begin{equation}
c_{s}^{2}(T)=\frac{\partial P}{\partial \varepsilon }=\frac{\sigma }{T}\frac{\partial T}{\partial \sigma }.
\label{cs2}
\end{equation}
For boost-invariant systems with cylindrical symmetry, Eqs. (\ref{hydTs}) may be put into a compact form. In this case: $v_r=\tanh \alpha$, where $\alpha$ is the transverse rapidity, and the longitudinal velocity is equal to $v_z=z/t$. With the help of auxiliary functions $a_\pm=\exp\left(\Phi\pm\alpha\right)$ and the $\Phi$ potential defined by the differential formula $ d\Phi = d\ln T / c_{s}=c_{s}d\ln \sigma$, the characteristic form of the hydrodynamic equations is introduced \cite{Baym:1983sr}
\begin{equation}
\frac{\partial }{\partial t}a_{\pm }\left( t,r\right) +\frac{v_{r}\pm c_{s}} {1\pm v_{r}\,c_{s}}\frac{\partial }{\partial r}\,a_{\pm }\left( t,r\right) + \frac{c_{s}}{1\pm v_{r}\,c_{s}}\left( \frac{v_{r}}{r}+\frac{1}{t}\right) a_{\pm }\left( t,r\right) = 0.
\label{eqnapm}
\end{equation}
Knowing the solutions of this equation, i.e., the functions $a_\pm$, we are able to calculate the transverse velocity and the $\Phi$ function from the expressions
\begin{equation}
v_r=\frac{a_+ - a_-}{a_+ + a_-}, \quad\quad \Phi = \frac{1}{2}\ln\left(a_+ a_-\right).
\label{vrPhiapm}
\end{equation}
\subsection{Temperature dependent sound velocity}\label{csofT}
We generalize the approach introduced in \cite{Baym:1983sr} by the use of the temperature dependent sound velocity in the numerical calculations \cite{Chojnacki:2004ec}. In order to evaluate the temperature from the auxiliary functions $a_\pm$ we first obtain $\Phi$ as a function of the temperature. We denote this function as $\Phi_T$. With the help of the inverse function, denoted by $T_\Phi$, we calculate the temperature from the formula
\begin{equation}
T\left(r,t\right)=T_\Phi\left(\frac{1}{2}\ln\left(a_+ a_-\right)\right).
\label{tempapm}
\end{equation}
In our numerical calculations we consider two forms of the sound velocity. The first one is obtained from the QCD lattice calculations \cite{Mohanty:2003va} and extrapolated to lower temperatures, while the second one is the analytic parameterization of these points. The latter form allowed us to calculate functions $\Phi_T$ and $T_\Phi$ analytically. 
\vspace*{-0.25cm}
\subsection{Initial conditions} \label{initcond}
We use the same method for constructing initial conditions as in \cite{Baym:1983sr}, i.e., a single function $a(r)$ is used to determine the initial forms of the functions $a_+$ and $a_-$, namely $a_+(t=t_0,r)=a(r), \, a_-(t=t_0,r)=a(-r)$.
Such procedure insures that a cylindrically symmetrical system has vanishing velocity field at $r=0$. In addition, we assume that the initial temperature profile is connected with the thickness function $T_A\left(r\right)$. We assume that the initially produced entropy density is proportional to the number of nucleons participating in the collision at the distance $r$ from the collision center, $\sigma(r) \sim T_A(r)$, and at high temperature our system may be considered as a system of massless particles. Since the entropy density of a massless gas is proportional to $T^3$ we find
\begin{equation}
T\left(r,t=t_0\right)=\hbox{const}\,\,T_A^\frac{1}{3}\left(r\right).
\label{initT}
\end{equation}
The final form of the function $a(r)$ that satisfies initial conditions (\ref{initvr}) and (\ref{initT}) reads
\begin{equation}
a(r,t=t_0) = a_T(r) \frac{\sqrt{1 + v^{\,0}_r}}{\sqrt{1 - v^{\,0}_r} }, \quad
a_T(r) = \hbox{exp}\left[\Phi_T\left( \hbox{const} \,\, T_A^\frac{1}{3}(|r|)\right)\right].
\label{inita2}
\end{equation}
\vspace*{-0.75cm}
\section{Results}\label{results}
\begin{figure}[!ht]
\vspace*{-0.25cm}
\begin{center}
\subfigure{\includegraphics[angle=0,width=0.31\textwidth]{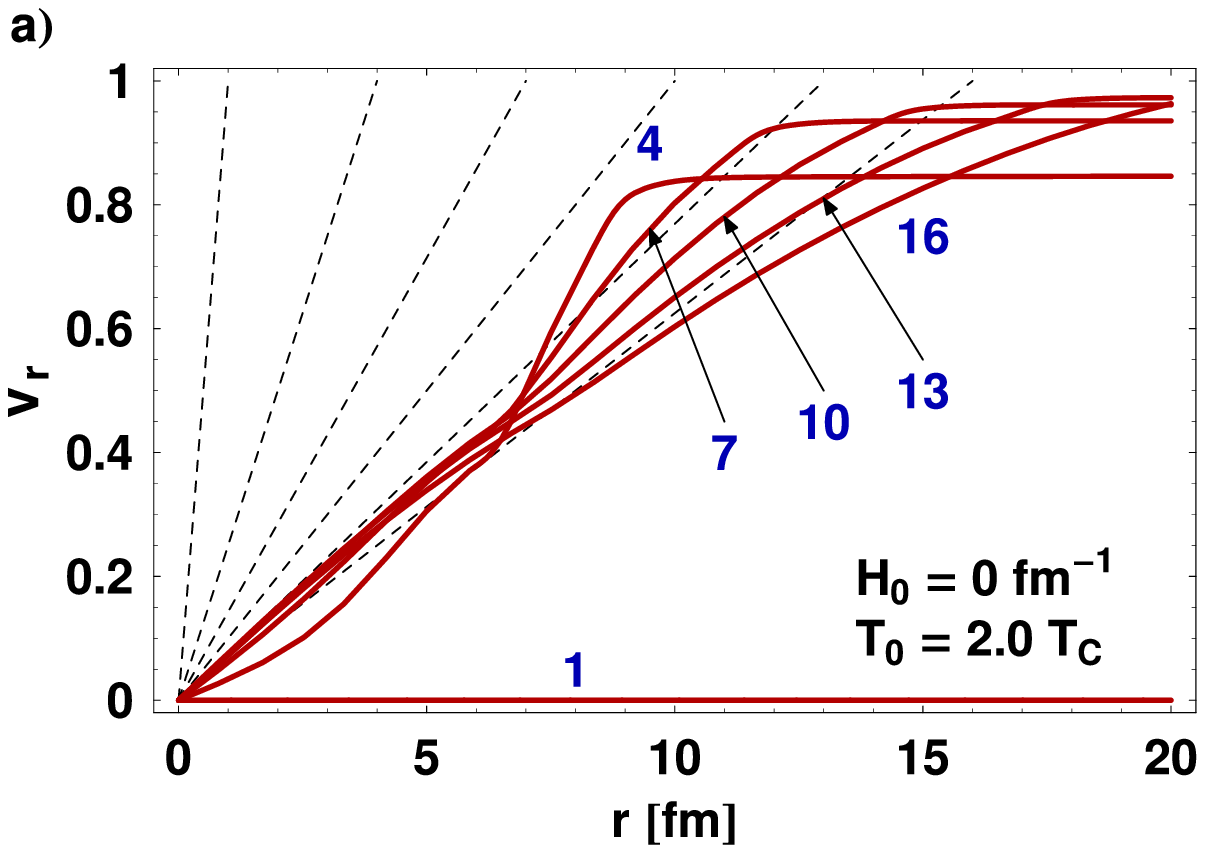}} \hspace*{2mm}
\subfigure{\includegraphics[angle=0,width=0.31\textwidth]{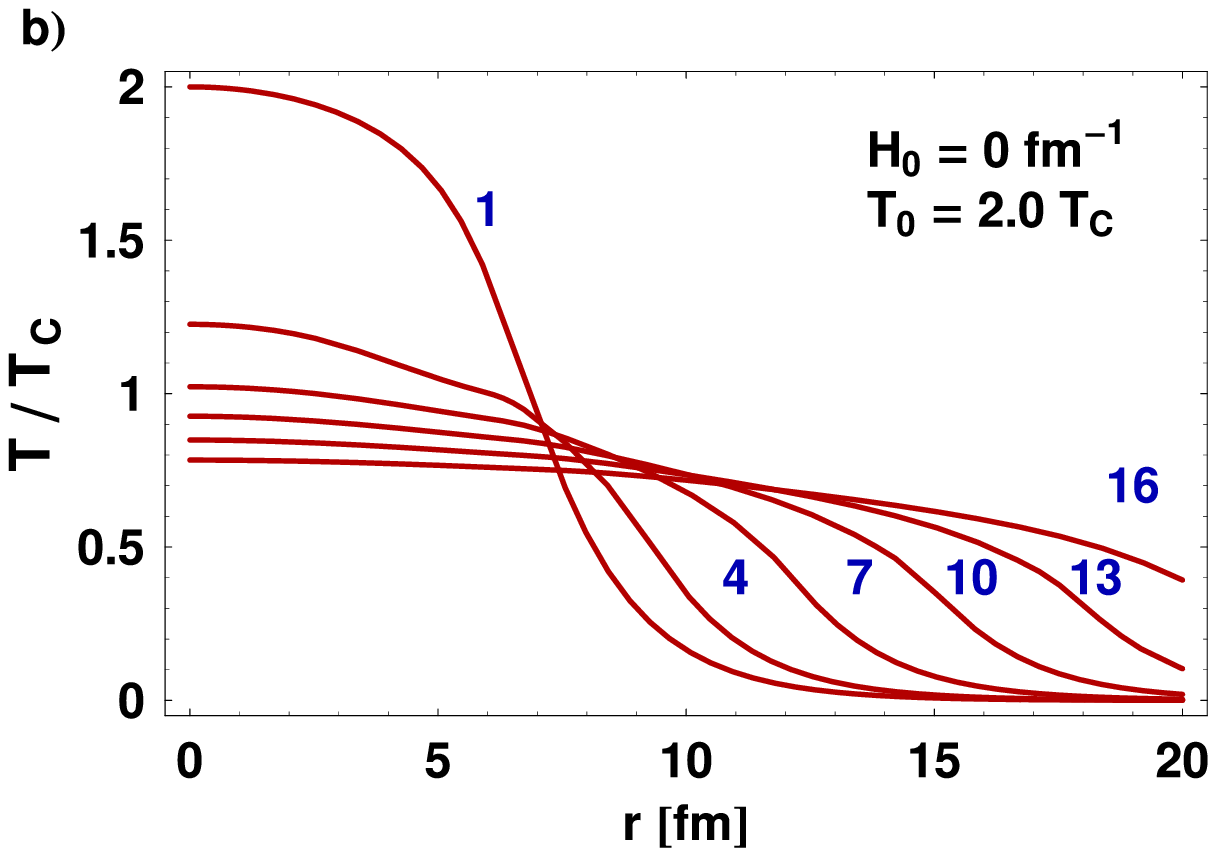}} \hspace*{2mm}
\subfigure{\includegraphics[angle=0,width=0.30\textwidth]{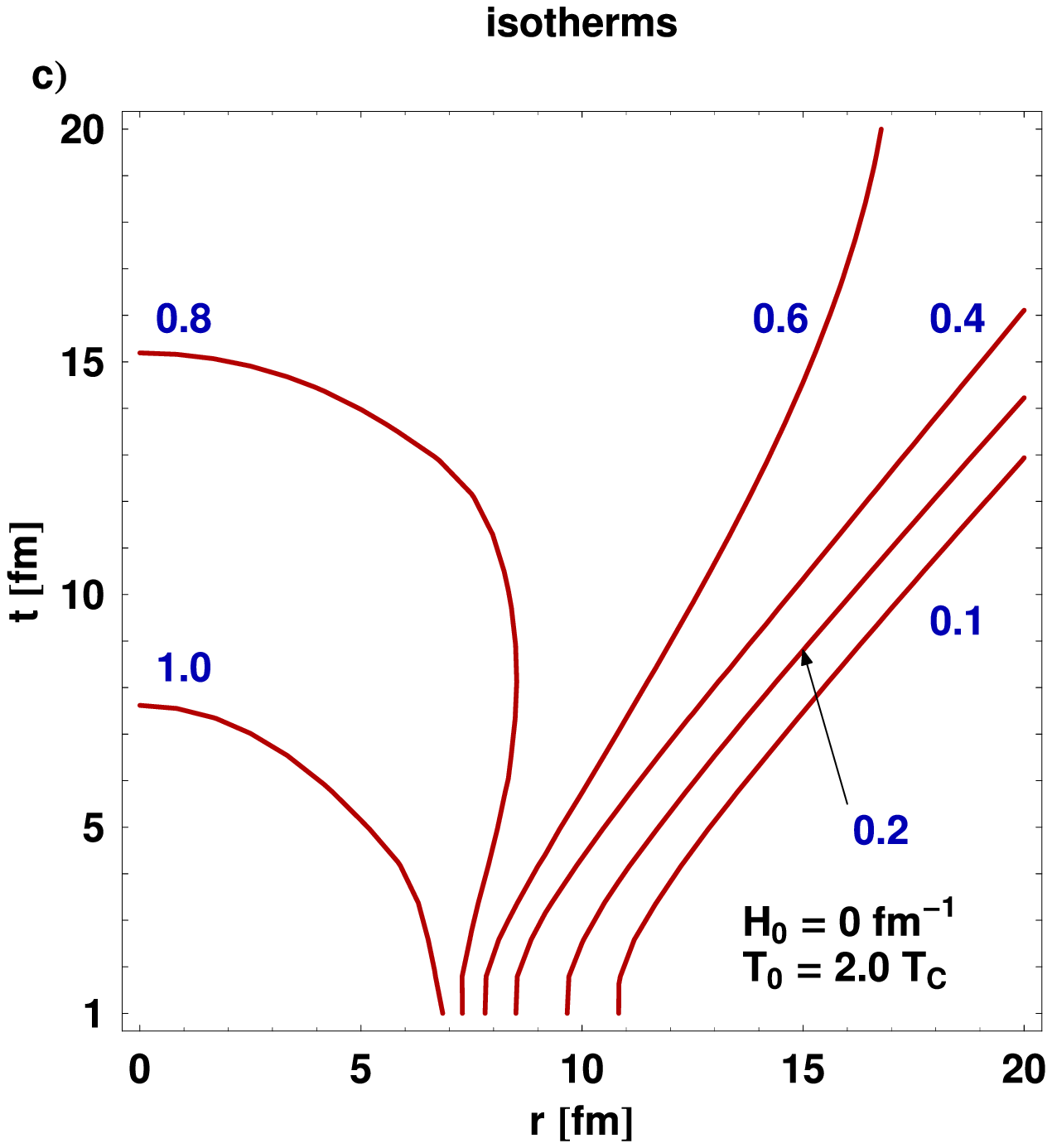}}\\
\subfigure{\includegraphics[angle=0,width=0.31\textwidth]{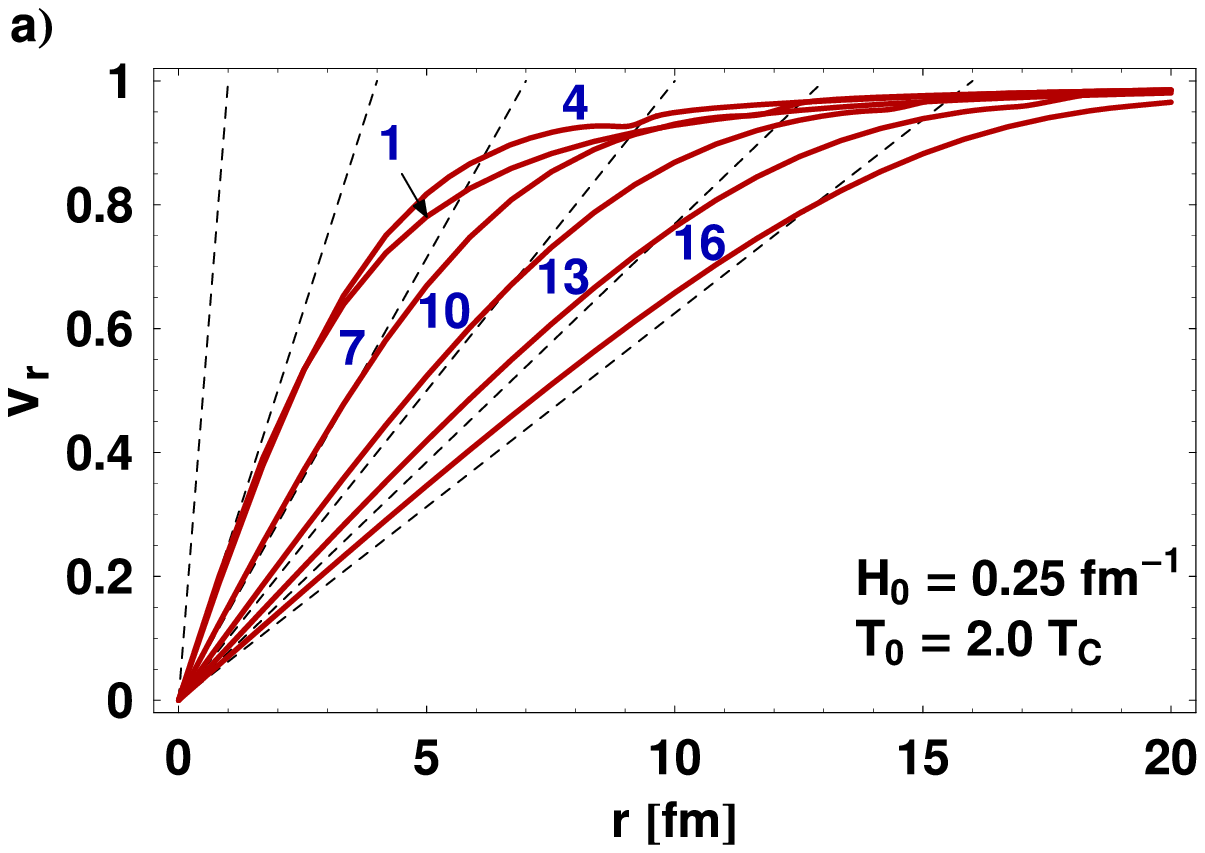}} \hspace*{2mm}
\subfigure{\includegraphics[angle=0,width=0.31\textwidth]{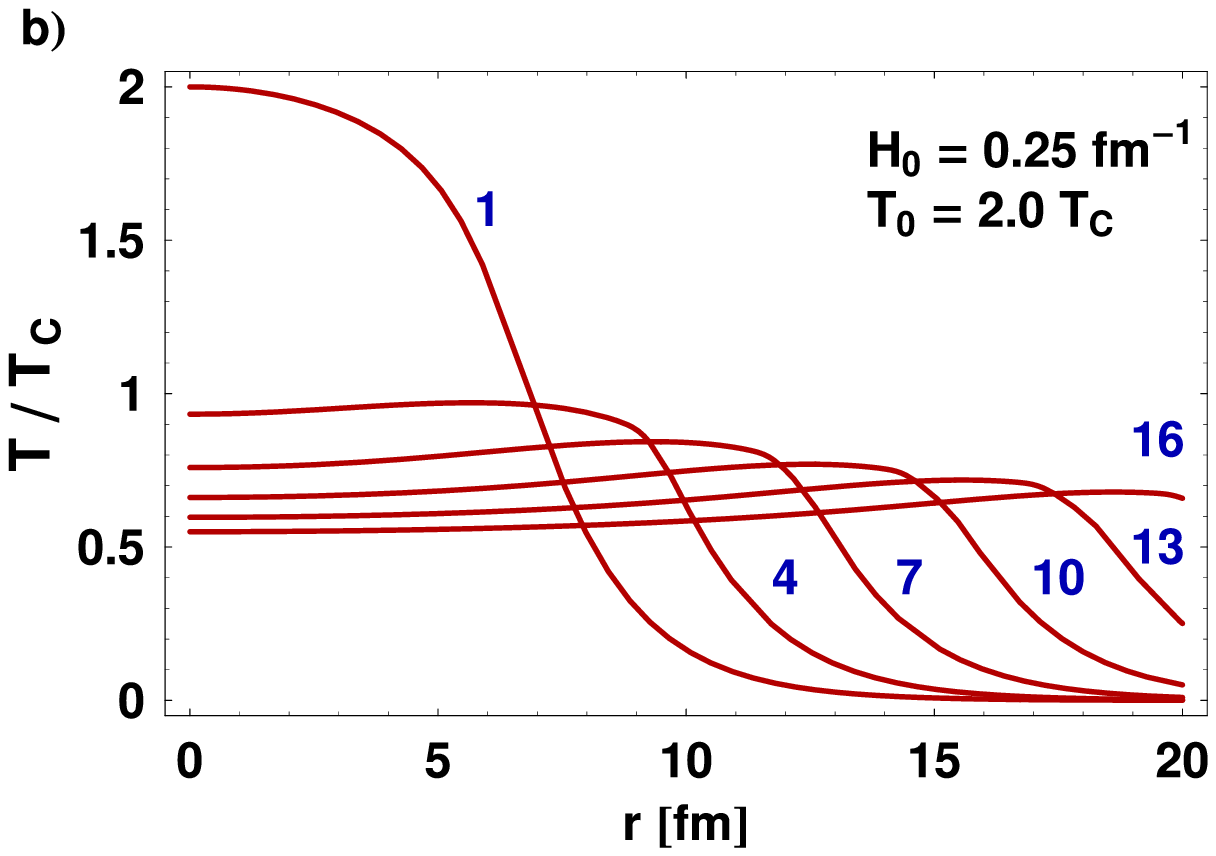}} \hspace*{2mm}
\subfigure{\includegraphics[angle=0,width=0.30\textwidth]{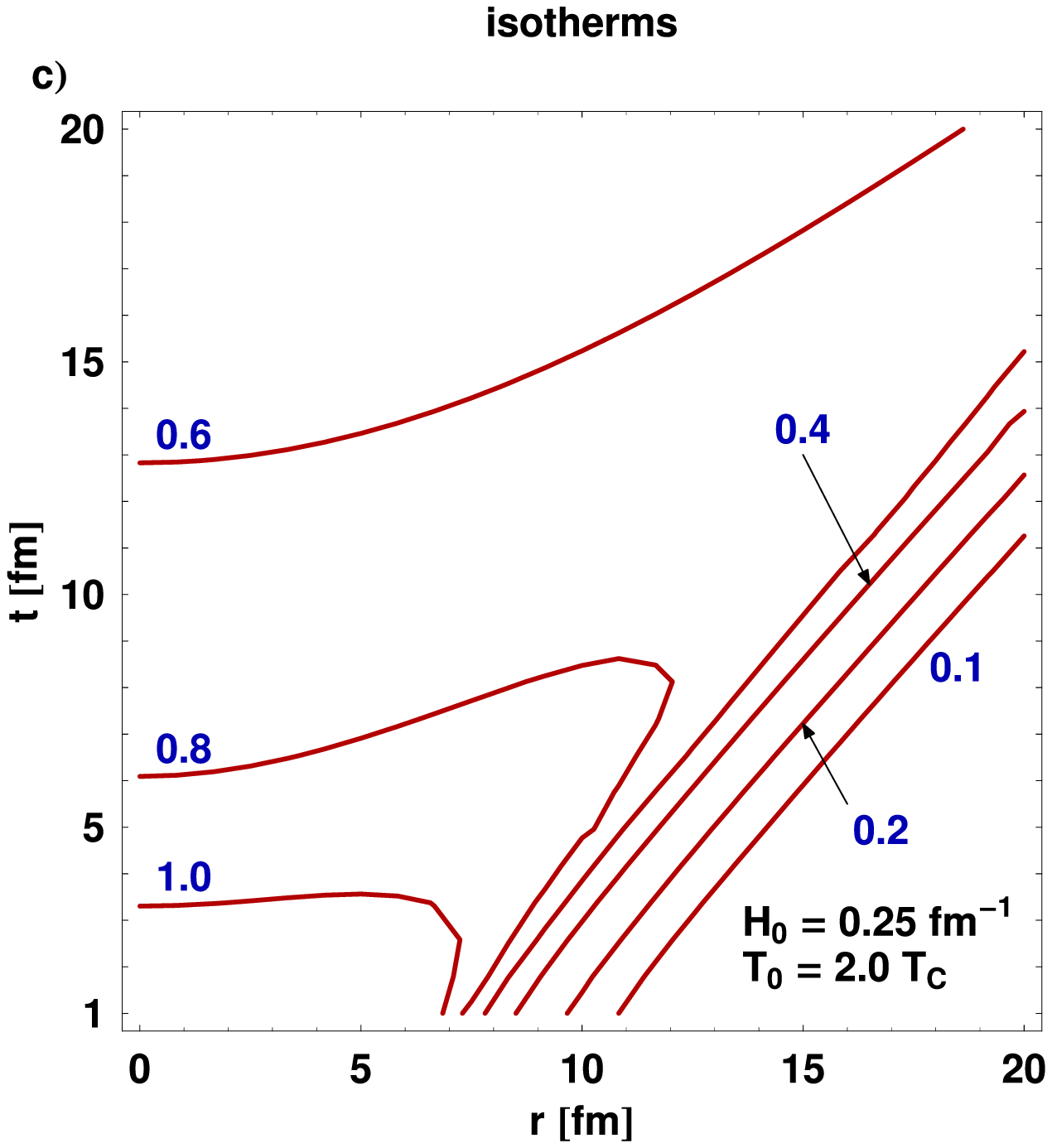}}
\end{center}
\vspace*{-1cm}
\caption{{\small Two cases of the hydrodynamic expansion of matter. The first case corresponds to matter being initially at rest, $H_0=0$, and the second case includes the pre-equilibrium flow characterized by the velocity profile (\ref{initvr}) with $H_0=0.25 \hbox{ fm}^{-1}$. The initial central temperature $T_0$ equals $2\,T_c$ ($T_c$ = 170 MeV), while the sound velocity is taken from the QCD lattice calculations. The parts {\bf a)} show the transverse velocity as functions of the distance from the center for 6 different values of time, $t_i$\,=\,1,\,4,\,7,\,10,\,13 and 16 fm. The dashed thin lines describe the ideal Hubble-like profiles of the form $r/t_i \, (r < t_i)$. The parts {\bf b)} show the temperature profiles in $r$, whereas the parts {\bf c)} show the isotherms in the $t-r$ space.  In this case, the labels at the curves denote the values of the temperature in units of the critical temperature.}}
\label{Fig_LAT}
\end{figure}

Our numerical studies show that the absence of the initial pre-equillibrium flow, the case $H_0=0$, unables formation of the scaling solutions during short evolution times, $t < 10 $ fm (see the upper parts of Fig. \ref{Fig_LAT}). In the case where the pre-equilibrium flow is included in the initial conditions we observe that the velocity profiles may approach the asymptotic scaling solutions after about 7 fm (see the lower parts of Fig. \ref{Fig_LAT}). The temperature profiles obtained in all considered cases show similar behavior; the system cools fast until it reaches the critical temperature and further cooling is slowed down. However, in the case of the non-zero initial flow the decrease of the temperature is faster and after some time the inner part of the system has lower temperature then the edge. This leads to the hyperbolic shape of the isotherms, as shown in the lower part of Fig. \ref{Fig_LAT} b). We conclude that this type of behavior gives support for the parameterizations used in the Blast-Wave \cite{Schnedermann:1993ws}, Buda-Lund \cite{Csorgo:1999sj}, and Cracow \cite{Broniowski:2001we} models.
\vfill\eject
\end{document}